# RHEOLOGICAL INTERPRETATION OF RAYLEIGH DAMPING

J.F.Semblat

*Laboratoire Central des Ponts et Chaussées, Eng. Modelling Div.,
58, bd Lefèbvre, 75732 Paris Cedex 15, France
email : semblat@lcpc.fr*

## 1. INTRODUCTION

Damping is defined through various terms [2] such as energy loss per cycle (for cyclic tests), logarithmic decrement (for vibration tests), complex modulus, rise-time or spectrum ratio (for wave propagation analysis)... For numerical modeling purposes, another type of damping is frequently used : it is called Rayleigh damping. It is a very convenient way of accounting for damping in numerical models, although the physical or rheological meaning of this approach is not clear. After the definition of Rayleigh damping, we propose a rheological interpretation of Rayleigh damping.

## 2. RAYLEIGH DAMPING

Rayleigh damping is a classical method to build easily the damping matrix $\boldsymbol{C}$ of a numerical model [5,7] under the following form :

$$\boldsymbol{C} = a_0 \boldsymbol{M} + a_1 \boldsymbol{K} \qquad (1)$$

where $\boldsymbol{M}$ and $\boldsymbol{K}$ are the mass and stiffness matrices respectively. It is then called *Rayleigh damping matrix*. $\boldsymbol{C}$ is the sum of two terms : one is proportional to mass matrix, the other to stiffness matrix.

A more general form was proposed by Caughey [3]. The original form (equation (1)) is very convenient as it can be easily computed. Furthermore, for modal approaches, the *Rayleigh (or Caughey) damping matrix* is diagonal in the real modes base [4,8]. Damping is therefore called proportional or classical. In case of non proportional damping, the complex modes have to be computed (in order to uncouple the modal equations).

Considering Rayleigh damping [4], the loss factor $\eta$ can be written as follows :

$$\eta = 2\xi = \frac{a_0}{\omega} + a_1 \omega \qquad (2)$$

where $\omega$ is the circular frequency and $\xi$ is the damping ratio.
Our purpose is to find out a rheological model having the same attenuation-frequency dependence as in the case of Rayleigh damping.

## 3. RHEOLOGICAL INTERPRETATION OF RAYLEIGH DAMPING

Considering the relationship between internal friction and frequency for Rayleigh damping, it is possible to build a rheological model involving the same attenuation-frequency dependence. For a linear viscoelastic rheological model of complex modulus $\boldsymbol{E}^* = \boldsymbol{E_R} + i.\boldsymbol{E_I}$ [2], expression of the quality factor $\boldsymbol{Q}$ is given in the fields of geophysics and acoustics as follows :

$$\boldsymbol{Q} = \frac{E_R}{E_I} \qquad (3)$$

For weak to moderate Rayleigh damping, there is a simple relation between the inverse of the quality factor $Q^{-1}$ and the damping ratio $\xi$ :

$$Q^{-1} \approx 2\xi \qquad (4)$$

For Rayleigh damping, the loss factor is infinite for zero and infinite frequencies. It clearly gives the behaviour of the model through instantaneous and long term responses. The rheological model perfectly meeting these requirements (attenuation-frequency dependence, instantaneous and long term effects) is a particular type of generalized Maxwell model. Figure (1) gives a schematic of the proposed model : it connects, in parallel, a classical Maxwell cell to a single dashpot. The generalized Maxwell model given in figure (1) can be defined through its complex modulus from which we easily derive the inverse of the quality factor $Q^{-1}$ which takes the same form than the loss factor of Rayleigh damping (expression (2)) : it is the sum of two terms, one proportional to frequency and one inversely proportional to frequency

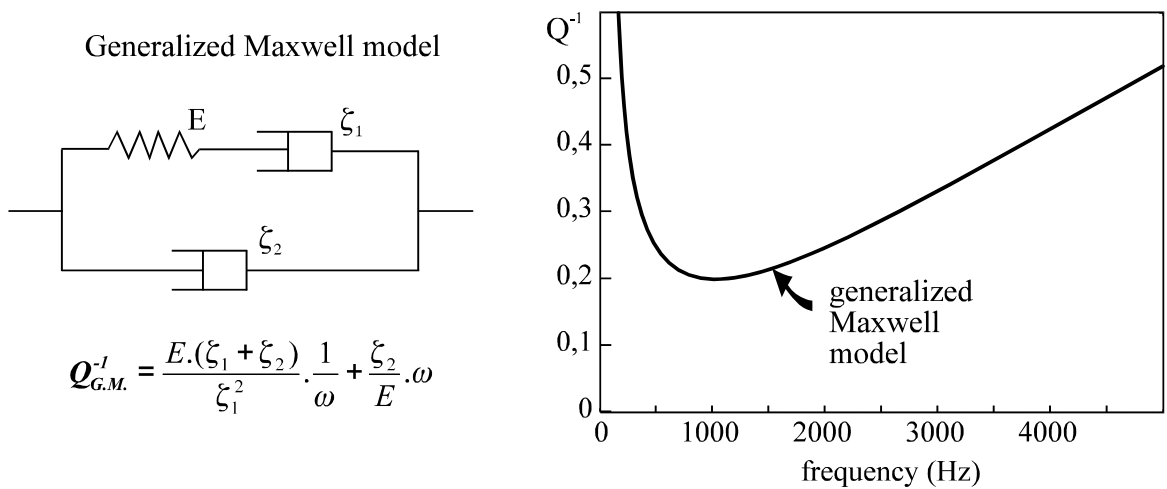

Fig. 1. Proposed generalized Maxwell model and corresponding attenuation curve

## 4. COMPARISON BETWEEN NUMERICAL AND ANALYTICAL RESULTS

### 4.1 Two different approaches

The coincidence between Rayleigh damping and the generalized Maxwell model is perfect considering internal friction (see equation (1) and figure (1) ). As equation (1) is only valid for moderate values of damping ratio $\xi$, there is a complete equivalence between both approaches since material velocity dispersion is moderate for such values of $\xi$. **Rayleigh damping and the generalized Maxwell model are equivalent for wave propagation purposes for small to moderate values of damping ratio**. A one-dimensional propagation test is then proposed to demonstrate the coincidence for moderate values of damping ratio $\xi$ and quantify the discrepancy for higher $\xi$ values. Rayleigh damping is investigated through a finite element modeling of the problem, whereas results for generalized Maxwell model are drawn from an analytical description (complex wavenumber derived from complex modulus [1,9,10,11]).

The numerical modeling is performed with CESAR-LCPC : the finite element program developed at LCPC and dedicated to civil engineering problems [6]. We use a one-dimensional mesh with linear quadrilateral elements and the finite element program performs a direct time integration. Rayleigh damping is involved considering expression (1). Analytical approach is based on the one-dimensional wave equation in which the material has viscoelastic properties corresponding to the generalized Maxwell model of figure (1). Harmonic solutions of the wave propagation problem in the frequency domain are found first and synthetized afterwards into the time domain [1,9,10,11].

*4.2 The problem and its parameters*

A one-dimensional wave propagation problem is chosen to compare numerical and analytical results. The point is to link numerical parameters of Rayleigh damping, that is $a_0$ and $a_1$ (see eq.(1)), to mechanical parameters of generalized Maxwell model, that is $E$, $\zeta_1$ and $\zeta_2$ (see figure (1)).

Considering eq.(2) and figure (1), these parameters can be easily related under the following form :

$$\begin{cases} a_0 = \dfrac{E.(\zeta_1 + \zeta_2)}{\zeta_1^2} \\ a_1 = \dfrac{\zeta_2}{E} \end{cases} \quad (5)$$

Expression (5) relates the Rayleigh coefficients ($a_0$ and $a_1$) to the behaviour parameters ($E$, $\zeta_1$ and $\zeta_2$) making experimental determination of Rayleigh coefficients much easier. For this numerical test, the applied loading is a sine-shaped single pulse ($\omega=10000$ rad.s$^{-1}$). Young modulus is $E=300$ MPa. For finite element model, the time step is $\Delta t=10^{-5}$ s and the elements dimension is chosen to minimize numerical dispersion effects.

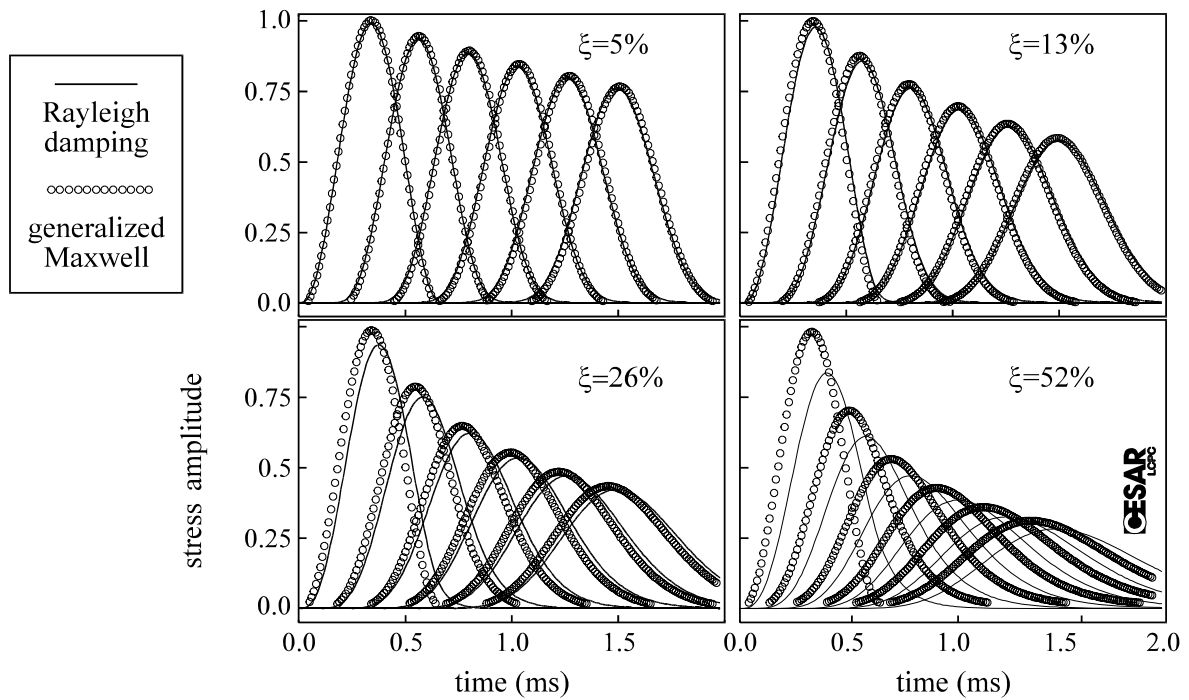

Fig. 2. Comparison between numerical results (Rayleigh damping)
and analytical results (generalized Maxwell model)

The results presented in fig.(2) correspond to six different distances from the source of excitation : 0, 0.1, 0.2, 0.3, 0.4 and 0.5m. Values of Rayleigh coefficients range from $a_0$=40 and $a_1$=$10^{-5}$ (first diagram) to $a_0$=400 and $a_1$=$10^{-4}$ (last diagram). For such values of Rayleigh coefficients, the corresponding mechanical parameters of the generalized Maxwell model are estimated using eq.(5) : $\zeta_1$=7.5.$10^6$ Pa.s, $\zeta_2$=3000 Pa.s (for first diagram).

The main conclusions draught from these curves are the following :
- for moderate values of damping coefficient (fig.(2)), numerical and analytical results perfectly coincide in terms of amplitude reduction and phase delays. It gives a good illustration of the theoretical link between Rayleigh damping and the generalized Maxwell model proposed in figure (1)
- for higher values of damping ratio ($\xi$>25%), attenuation is stronger for Rayleigh damping approach and dispersive phenomena are different in both cases

## 5. CONCLUSION

A rheological model is proposed to be related to classical Rayleigh damping : it is a generalized Maxwell model with three parameters (fig.(1)). For moderate damping ($\xi$<25%), this model perfectly coincide with Rayleigh damping approach since internal friction has the same expression in both cases and dispersive phenomena are negligible. This is illustrated by finite element (Rayleigh damping) and analytical (generalized Maxwell model) results in a simple one-dimensional case.

## ACKNOWLEDGEMENTS

The author is indebted to Pr O. Coussy for his pertinent comments on the preliminary version of the manuscript.